\documentclass[final,1p,times]{elsarticle} 
\usepackage{graphicx} 
\usepackage{amssymb} 
\usepackage{amsthm} 
\usepackage{lineno} 

\newcommand{\jpsi}{\rm J/$\psi$}
\newcommand{\sigmaabs}{$\sigma_{abs}^{\rm J/\psi}$}

\journal{Nuclear Physics A} 
\begin{document} 

\begin{frontmatter} 


\title{\rm J/$\psi$ production in \mbox{p-A} and \mbox{A-A} collisions at fixed 
target experiments}

\author{Roberta Arnaldi$^{a}$ for the NA60 Collaboration}

\address[a]{INFN Torino, 
Via P. Giuria 1,
Torino, I-10125, Italy}

\begin{abstract} 
Charmonia suppression is one of the main signatures for the formation of a  
deconfined medium. However, also nuclear effects, not related to the
production of a hot medium, can affect the \jpsi\ yield. 
The determination, from the study of \mbox{p-A} collisions, 
of the \jpsi\ behaviour in nuclear matter  
is, therefore, extremely important to correctly quantify the amount of charmonia 
suppression induced by the deconfined medium. 
In this paper the new NA60 results collected at 158 GeV incident energy, i.e. under the same
kinematical conditions as the In-In (NA60) and Pb-Pb (NA50) data, are presented
and compared with \mbox{p-A} measurements from other fixed target 
experiments. 
Results on \mbox{A-A} collisions are also reviewed taking into
account the new available information on the influence of cold nuclear matter 
on the \jpsi\ production yield.
Finally, results on the J/$\psi$ polarization are shown for \mbox{p-A} and
\mbox{A-A} collisions.

\end{abstract} 

\end{frontmatter} 



\section{Introduction}

\jpsi\ suppression in \mbox{A-A} collisions is considered to be one of the main
signatures for the phase transition from an hadronic to a deconfined 
medium~\cite{Mat86}. 
However, the \jpsi\ production can be affected also by cold nuclear matter
effects, as, for example, the final state absorption of the $c\bar{c}$ pair in the nuclear medium, the
parton shadowing or the initial and final state energy loss.
The amplitude of these effects, not related to the formation of a deconfined 
medium, is determined from \mbox{p-A} collisions and then 
extrapolated to \mbox{A-A} interactions to be compared with the measured 
\jpsi\ yield.  

At SPS energies, cold nuclear matter effects are usually parameterized, in 
the frame of the Glauber model, by fitting the A-dependence of the \jpsi\ 
production cross section per nucleon-nucleon collision. These nuclear effects 
are quantified through the value of \sigmaabs\ extracted, up to now, from the
NA50 data at 400/450 GeV (\sigmaabs\ = 4.2 $\pm$ 0.5 mb)~\cite{Ale06}.
Another way to parameterize cold nuclear matter effects
is based on a fit to the A-dependence of the \jpsi\ production cross section using a $A^{\alpha}$ 
power law. 
A value of $\alpha$ different from 
unity indicates how much the \jpsi\ yield is modified by the nuclear medium.
It should be noted that both $\alpha$ and \sigmaabs\ are effective quantities, 
since they represent the amount of cold nuclear matter effects reducing the 
\jpsi\ yield, but they do not allow to disentangle the different contributions (e.g.
shadowing, nuclear absorption) playing a role in this reduction.
Nuclear effects, evaluated from \mbox{p-A} data at 400/450 GeV, are
then extrapolated to \mbox{A-A} collisions at a lower energy (158A GeV) assuming
in both cases a scaling with $\rm{L}$, the mean thickness of
nuclear matter seen by the $c\bar{c}$ pair in its way out through the nucleus,
and imposing \sigmaabs\ to be energy independent. 
The expected \jpsi\ yield is then 
compared with the measured one, as a function of centrality. 
Following the aforementioned approach, a further suppression (the so called 
``anomalous'' \jpsi\ suppression), exceeding cold nuclear matter effects,    
is indeed observed at SPS in \mbox{Pb-Pb}~\cite{Ale05} and 
\mbox{In-In}~\cite{Arn07} collisions.
It is clear that to correctly quantify the amount of suppression due to the
formation of a hot medium, the \jpsi\ 
behaviour in the normal nuclear matter must be determined with high precision.

Several results on \jpsi\ production in \mbox{p-A} collisions at fixed target 
experiments are available, covering different kinematic and 
energy domains.
HERA-B at HERA, for example, has studied \mbox{p-Cu, Ti, W} reactions at 920 GeV~\cite{Abt09}, 
E866 at FNAL \mbox{p-Be, Fe, W} collisions
at 800 GeV~\cite{Lei00}, NA50 at SPS proton induced 
collisions on  several nuclei (Be, Al, Cu, Ag, W, Pb) at 400/450
GeV~\cite{Ale06} and finally NA3 at SPS \mbox{p-H$_{2}$, Pt} collisions
at 200 GeV~\cite{Bad83}.
It is important to note that none of these existing results has been obtained 
under the same conditions, in terms of energy and kinematic domain, as the SPS 
\mbox{A-A} collisions collected at 158A GeV. 
This implies that several assumptions (as the energy independence of \sigmaabs\ 
) had to be done, in order to extrapolate cold nuclear matter effects from
\mbox{p-A} to \mbox{A-A} collisions.

\section{NA60 \mbox{p-A} results}

For the first time the NA60 experiment has studied the \jpsi\ production in 
\mbox{p-A} collisions at 158 GeV~\cite{Sco09}, in order to provide a reference 
collected under the same kinematic and energy conditions as the \mbox{A-A} data. 

The experimental apparatus of NA60 is based on a traditional muon spectrometer 
coupled with a vertex telescope made of Si pixel planes, close to the target 
region. The matching, based on the tracks coordinates and momentum between the 
tracks reconstructed in the two spectrometers, allows an accurate measurement of 
the muon kinematics and therefore an improvement in the quality of the results 
with respect to previous experiments. 
For details on the apparatus see~\cite{Usa05}.
The target system of NA60, during the \mbox{p-A} data taking, was based on 7 
different targets (Be, Al, Cu, In, W, Pb, U) simultanously exposed to the beam. 
NA60 has also collected data at 400 GeV, with the same experimental setup as
the one of the 158 GeV data taking period, to be used as a cross-check of the 
NA50 results taken at the same energy.

\subsection{$\rm J/\psi$ kinematical distributions}

\begin{figure}[ht]
\centering
\includegraphics[scale=0.33]{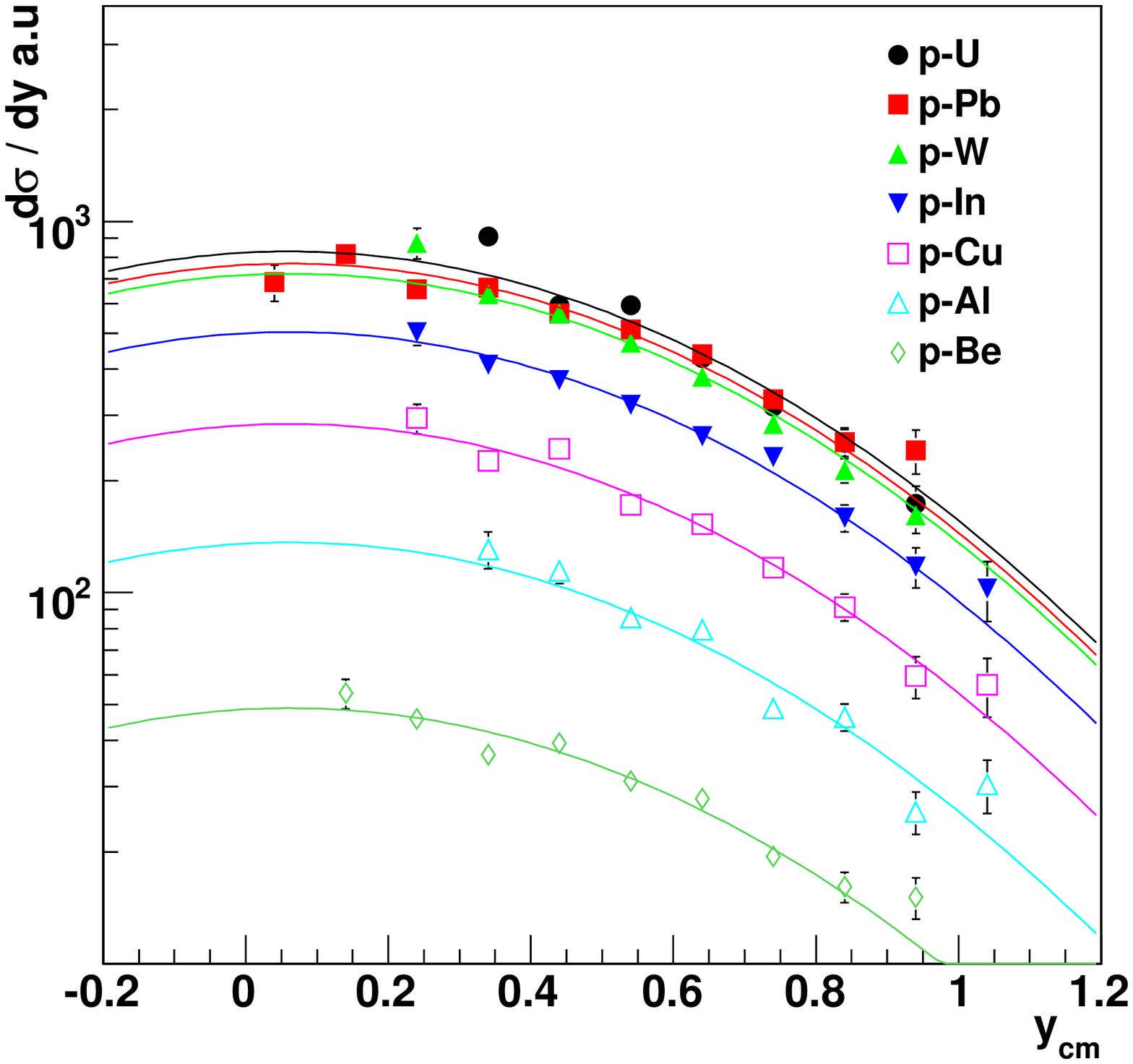}
\includegraphics[scale=0.33]{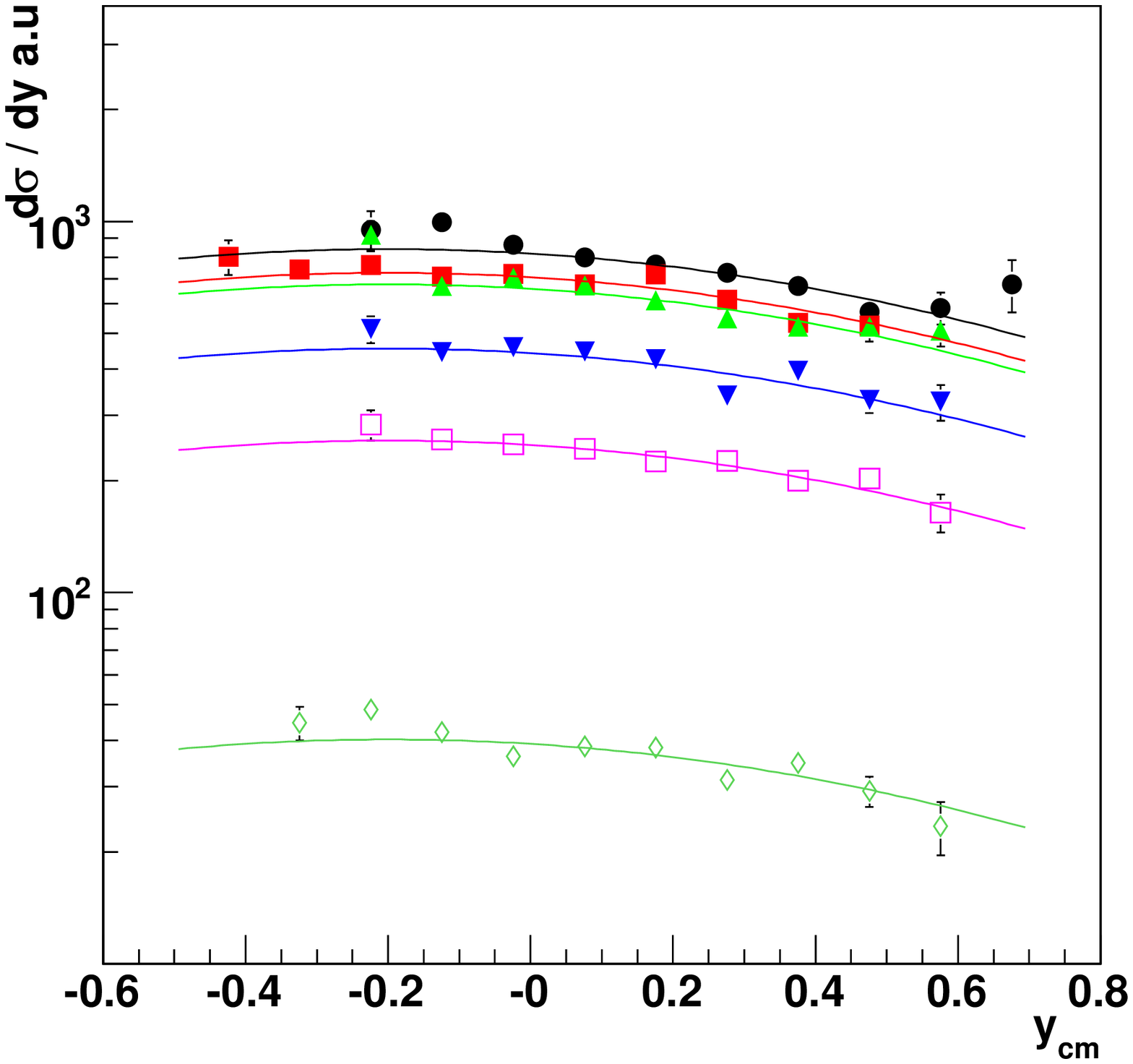}
\caption[]{\jpsi\ acceptance corrected rapidity distributions at 158 GeV (left)
and 400 GeV (right).}
\label{fig:1}
\end{figure}

The investigation of the \jpsi\ kinematical distributions may help to obtain
further insights in the understanding of the charmonium production and initial
state effects. 
In Fig.\ref{fig:1} the rapidity ($y_{\rm{CM}}$) 
distributions corresponding to \mbox{p-A} collisions at 158 GeV and 400 
GeV are shown. A 1-D acceptance correction has been performed assuming
realistic transverse momentum distributions and a flat cos$\theta_{CS}$ shape (see
Sec.\ref{sec:pol} for this definition).
Within the narrow rapidity coverage of NA60 (one rapidity unit), the 
distributions can be reproduced by a Gaussian function. A simultaneous fit to 
the distributions corresponding to the different \mbox{p-A} collisions indicates
that,  
at 158 GeV, the data can be described by a gaussian centered at midrapidity ($\mu_{\rm y}=0.05 \pm 0.05$) 
with a width $\sigma_{\rm y} = 0.51 \pm 0.02$.
At 400 GeV, because of the wider distributions, the peak position is less 
constrained. In this case the fit is performed imposing the slightly negative 
mean value ($\mu_{y}=-0.2$) measured by NA50 with a high statistics data set at the 
same energy~\cite{Ale06}. 
As expected, because of the higher energy, a larger rapidity width 
($\sigma_{y} = 0.81 \pm 0.03$) is obtained, in agreement, within errors with the 
NA50 value ($\sigma_{y} \sim 0.85$).
Unfortunately, the precision of these results do not allow to investigate the 
negative rapidity shift observed by NA50, neither to confirm the smaller A-dependent shift of the center of the
$x_{F}$ (and therefore $y$) distributions observed by HERA-B~\cite{Abt09}.

\begin{figure}[ht]
\centering
\includegraphics[scale=0.32]{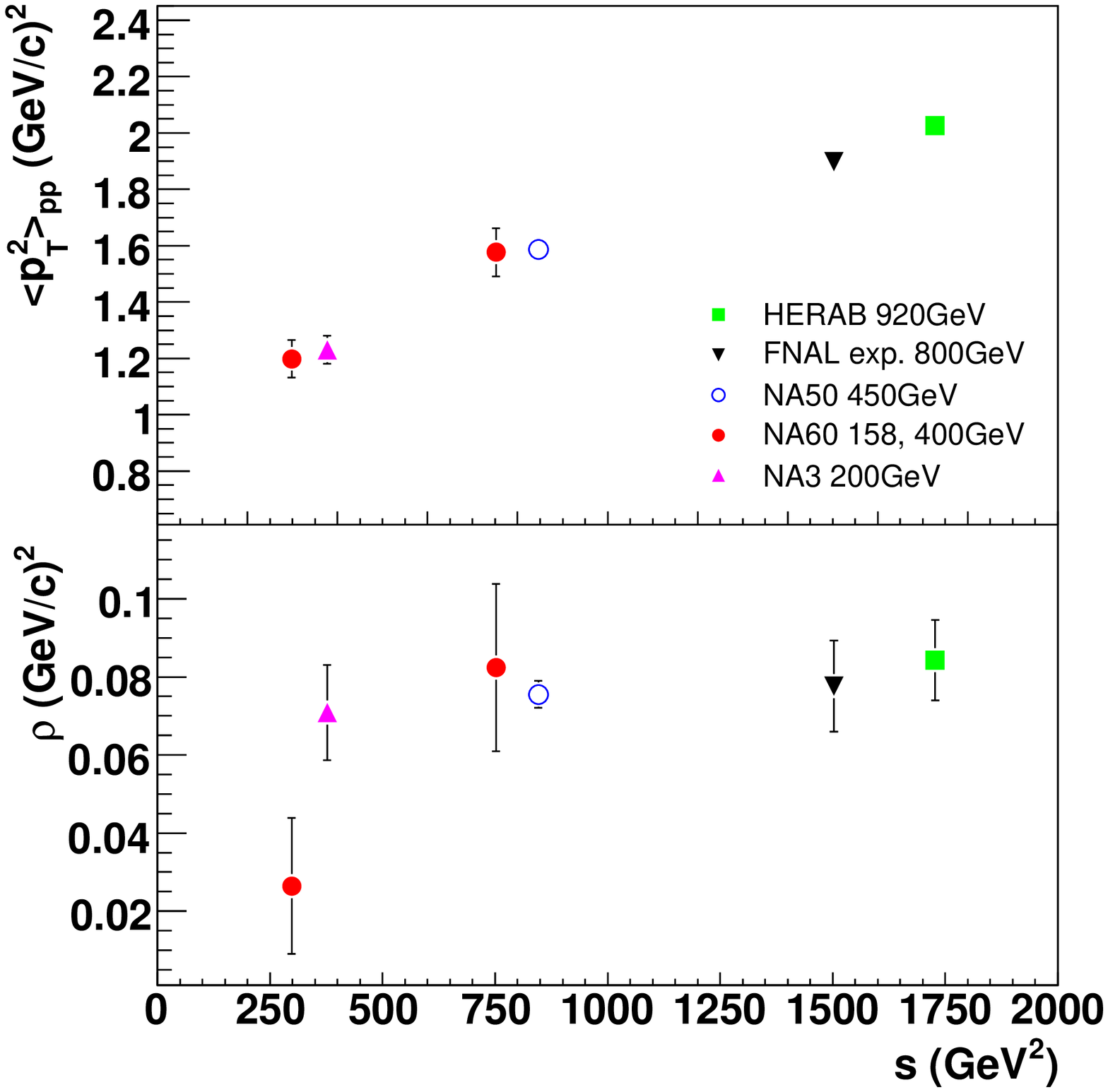}
\includegraphics[scale=0.33]{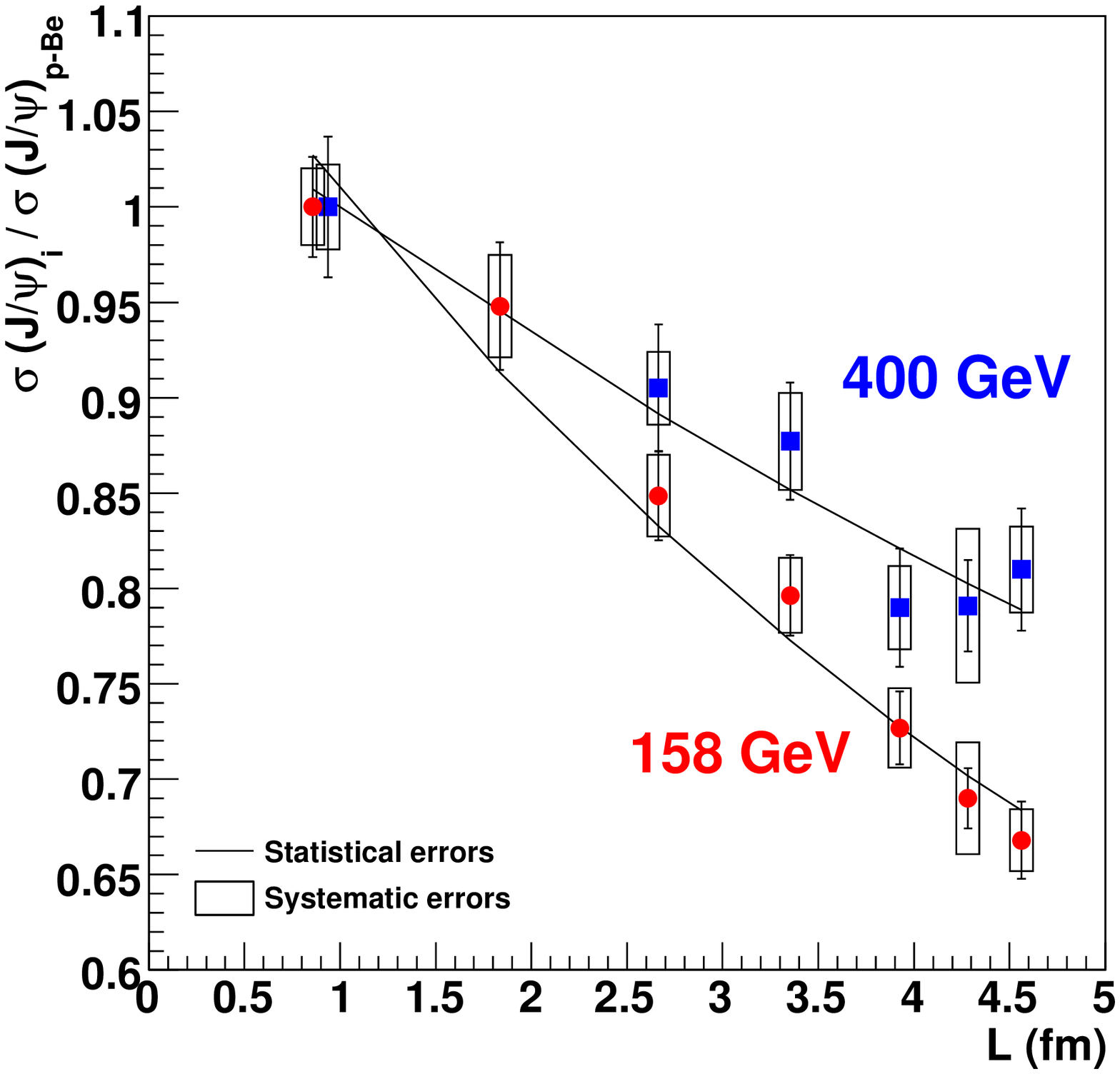}
\caption[]{Left: energy dependence of the $<p_{\rm T}^2>_{pp}$ (top) and $\rho$
(bottom) parameters. Right: \jpsi\ cross section ratios for \mbox{p-A}
collisions at 158 GeV (circles) and 400 GeV(squares), as a function of $\rm
{L}$.}
\label{fig:2}
\end{figure}

The \jpsi\ transverse momentum ($p_{\rm T}$) distributions have also been
studied. The spectra are corrected with an 1-D acceptance, obtained
assuming a flat $p_{\rm T}$ distribution and realistic $y$ and  
cos$\theta_{CS}$ shapes.
A broadening of the $p_{\rm T}$ distributions as a function of A is 
measured at both 158 and 400 GeV. This confirms previous 
observations from other experiments~\cite{Abt09,Bad83,Ale04,Kow94,Sch95,Ale97,
Gri01} and is usually interpreted as initial state 
multiple scattering of the incoming gluons (Cronin effect) before the 
hard scattering which will produce the $c\bar{c}$ pair.
The average $<p_{\rm T}^2>$ dependence on the mass number can be
parameterized with $<p_{\rm T}^2> = <p_{\rm T}^2>_{pp} + \rho
(A^{1/3}-1)$~\cite{Abt09}, where
$A^{1/3}-1$ is roughly proportional to the length $\rm L$. 
As shown in Fig.\ref{fig:2} (left) the $<p_{\rm T}^2>_{pp}$ values are compatible with a
linear growth with the square of the center of mass energy $\rm s$. 
On the other hand, the slope of the parameterization $\rho$ shows an almost 
energy independent pattern, apart from a hint of decrease at low energy, corresponding to the
NA60 158 GeV result.

 \subsection{Cold nuclear matter effects}

New NA60 results on the nuclear effects affecting the 
\jpsi\ at 158 GeV have been presented at this conference~\cite{Sco09}. 
Results, integrated over the full $p_{\rm T}$ range are given in the rapidity region, covered by all the subtargets, 
corresponding to $0.28 < y_{\rm CM} < 0.78$.
Nuclear effects are evaluated comparing the cross section ratio, 
$\sigma_{pA}^{\rm J/\psi} / \sigma_{pBe}^{\rm J/\psi}  $, between the target with
mass number A and the lightest one (Be).
By computing this ratio, the beam luminosity factors cancel out, 
apart from a small beam attenuation factor. 
However, since the sub-targets see the vertex telescope under a slightly 
different angle, the track reconstruction efficiencies do not completely cancel 
out. 
Therefore an accurate evaluation of such quantities and their time evolution has
been performed target by target, with high granularity down to the single
pixel level, and on a run-per-run basis.

\jpsi\ cross-section ratios are shown in Fig.\ref{fig:2} (right) 
as a function of $\rm L$. In the same figure, also results from a similar
analysis performed on the 400 GeV data sample are shown. In this case, the
results refer to the
kinematical region $-0.17< y_{\rm CM}< 0.33$, corresponding to the same 
rapidity range, in the laboratory frame, as the 158 GeV one.
Systematic errors include uncertainties on the target thickness, 
on the $\rm y$ distribution used in the acceptance calculation and on the reconstruction
efficiency. Only the fraction of systematic error, not in common to all the
points, is shown, since it is the one affecting the evaluation of the nuclear
effects.

Performing a Glauber fit to the data, or using the $\alpha$
parameterization, the following results are
obtained for the new NA60 \mbox{p-A} data: \sigmaabs\ = 7.6 $\pm$ 0.7 (stat.) $\pm$ 0.6
(syst.) mb ($\alpha$ = 0.882 $\pm$ 0.009 $\pm$ 0.008) 
at 158 GeV and \sigmaabs\ = 4.3 $\pm$ 0.8 (stat.) $\pm$ 0.6 (syst.) mb  ($\alpha$ = 0.927 
$\pm$ 0.013 $\pm$ 0.009) at 400 GeV.
It has to be stressed that the \sigmaabs\ result at 400 GeV is
smaller with respect to the one extracted from the 158 GeV, pointing to an
energy dependence of this quantity. Furthermore, the obtained value is in very 
good agreement with a previous result obtained by NA50 at the same 
energy~\cite{Ale06}.

\begin{figure}[ht]
\centering
\includegraphics[scale=0.335]{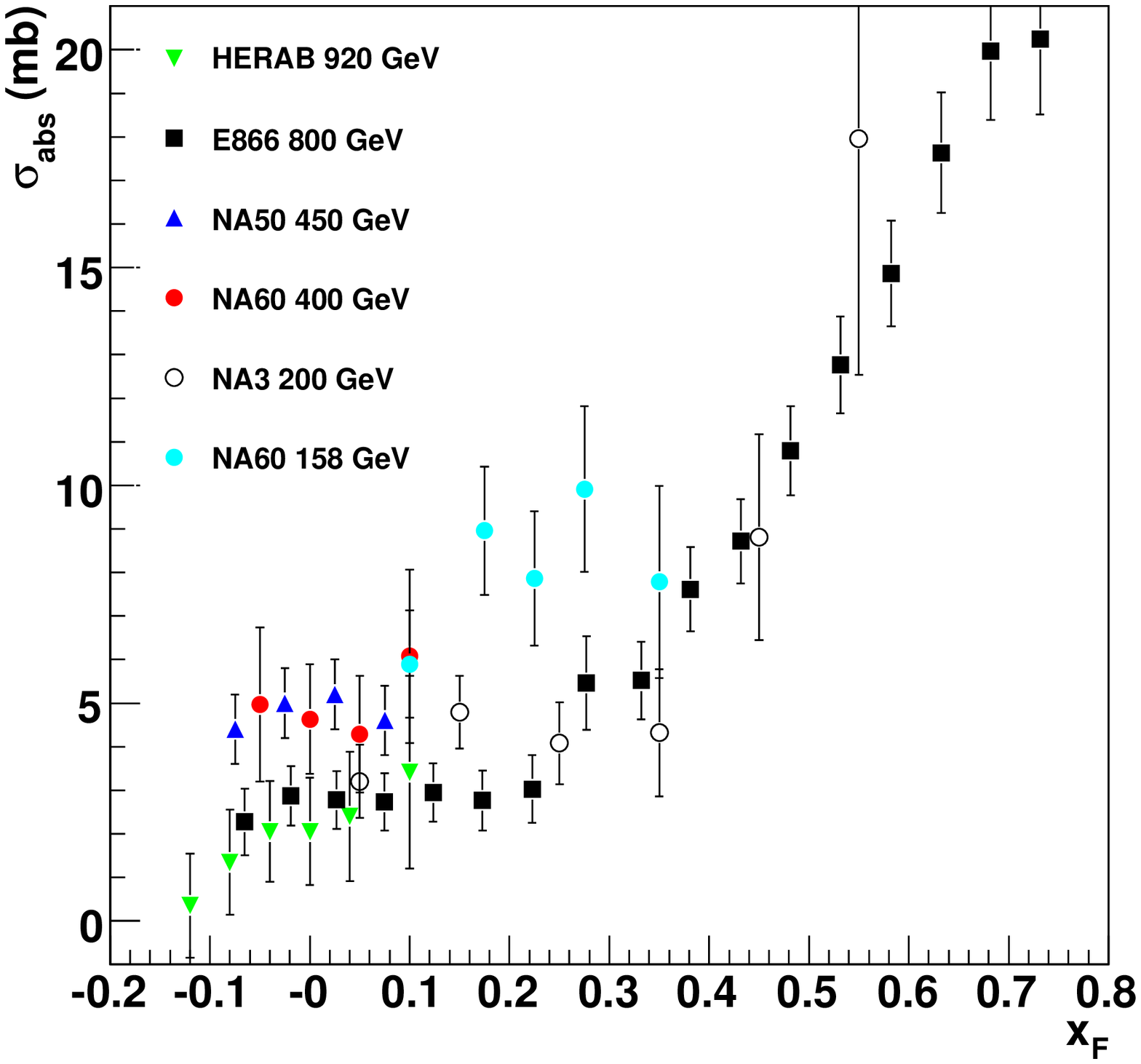}
\includegraphics[scale=0.33]{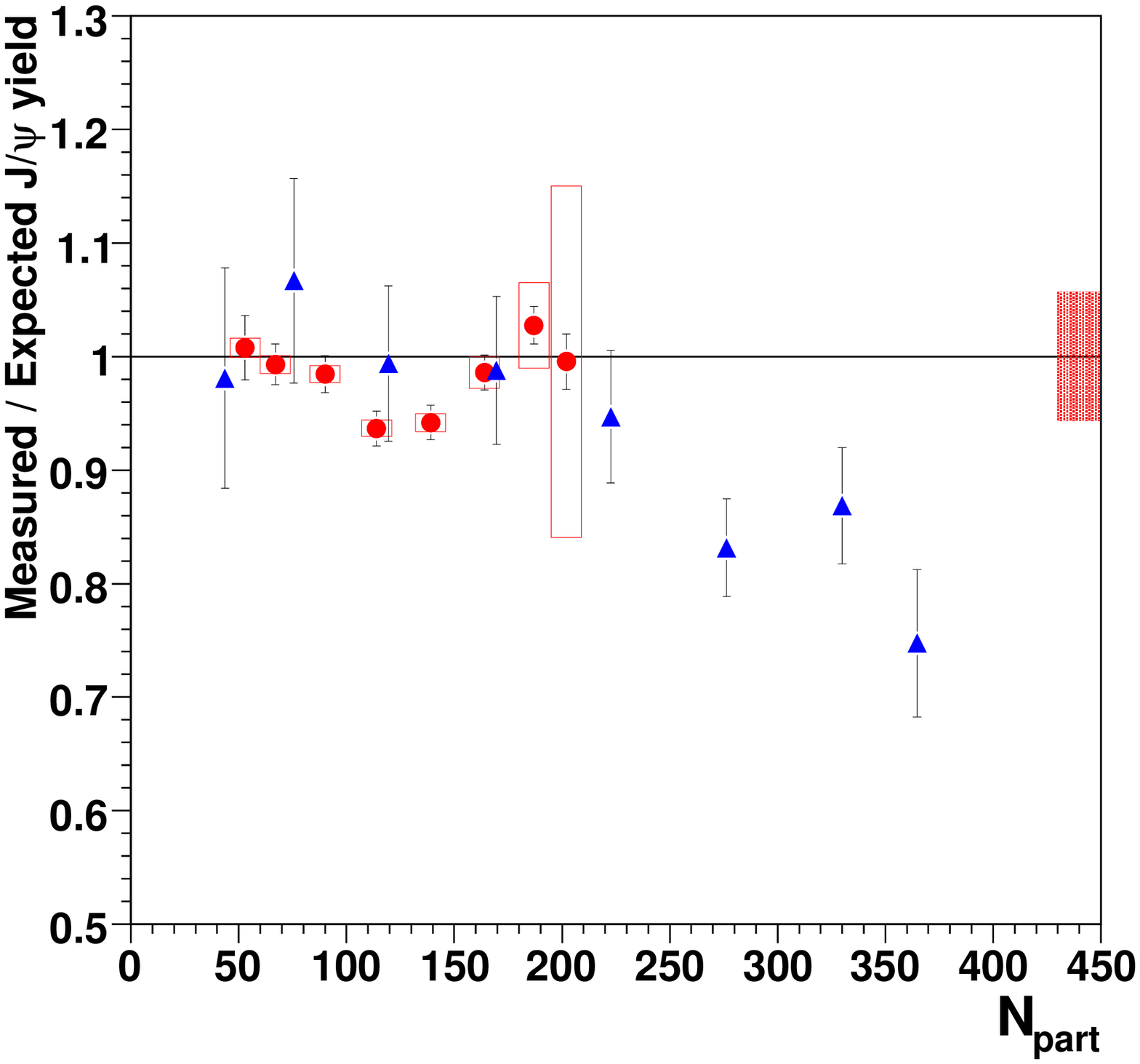}
\caption[]{Left: Compilation of \sigmaabs\ values versus $x_{\rm F}$. 
Right: \jpsi\ suppression pattern in \mbox{In-In} (circles) and
\mbox{Pb-Pb} (triangles). Boxes around the points correspond to the correlated
systematic errors, while the filled box on the right is the uncertainty
on the absolute normalization of the \mbox{In-In} points. A further global 
12\% error, not shown, is due to the uncertainty on \sigmaabs\ .}
\label{fig:3}
\end{figure}

The comparison with results from previous experiments (HERA-B, E866 and NA50) 
can be done in terms of the extracted $\alpha$~\cite{Sco09}  
or \sigmaabs\ values, as a function of the 
Feyman $x$ variable ($x_{\rm F}$). 
As shown in Fig.\ref{fig:3} (left), 
where all the available measurements are compared, a strong dependence of \sigmaabs\ on 
$x_{\rm F}$ and on the beam energy is clearly visible. Cold nuclear matter 
effects are stronger at high $x_{\rm F}$ and for a fixed  $x_{\rm F}$ value 
their importance increases while lowering the beam energy.
As shown in the figure, the new NA60 results at 400 GeV confirm the 
NA50 values obtained at a similar energy. On the other hand, the 158 GeV data 
seem to point to higher \sigmaabs\ values, with a hint for an increase in the
narrow explored $x_{\rm F}$ region. 
It is also worthwhile to note that older NA3 results~\cite{Bad83} on \jpsi\ 
production are in partial contradiction with these observations, showing 
lower \sigmaabs\ values, rather similar to 
those obtained with the higher energy data samples. 

Although not shown here, the results do not present any scaling with $x_{2}$, 
the fraction of nucleon momentum carried by the target colliding partons. 
Since $x_{2}$ is the variable driving the shadowing effects, the break of this 
scaling confirms the fact that parton shadowing can not be the only mechanism 
invoked to describe the data. 
It is, therefore, clear that the interpretation of the kinematical dependence of 
the cold nuclear matter
effects is extremely complicated since there are many competing mechanisms
affecting the \jpsi\ production and propagation into the nuclei.
A theoretical description of the cold nuclear matter effects over the full 
kinematic range is extremely complex and not yet
available~\cite{Vog00,Vog05,Bor03}. 

As discussed, the $\alpha$ or the \sigmaabs\ values are effective
quantities used to describe the convolution of all the cold nuclear effects not related to the
formation of a deconfined medium. 
At this conference, a first attempt to take explicitly into account the parton 
shadowing contribution to the NA60 data has been proposed. It must be noted that at SPS energies, the 
charmonia production explores a range 
of $x$ (the fraction of the nucleon momentum carried by the parton) 
corresponding to the antishadowing region, where parton densities 
in the nuclei are enhanced with respect to those of the free nucleons.
The antishadowing, evaluated with the 
EKS98\cite{Esk99} parameterization for the nuclear modification of 
the parton distribution functions (PDF), causes an enhancement of the charmonium
production cross section per nucleon-nucleon collision. Therefore, if this
initial-state 
contribution is explicitly taken into account, a larger \sigmaabs\ value is needed to
describe the measured data ($\sigma_{abs}^{\rm J/\psi}$(158GeV) = $9.3 \pm 0.7 \pm
0.7$ mb and $\sigma_{abs}^{\rm J/\psi}$(400GeV) = $6.0 \pm 0.9 \pm 0.7$ mb).
Results depend on the adopted parameterization of the PDF nuclear modifications. 
Slightly higher \sigmaabs\ values ($\sim 5-10$\%), for example, are obtained if
the EPS08~\cite{Eps08} parameterization is used.

\section{Anomalous $\rm J/\psi$ suppression in \mbox{In-In} and \mbox{Pb-Pb} collisions}

As previously discussed, the cold nuclear matter reference used up to now for
\mbox{A-A} collisions by 
NA50~\cite{Ale05} and NA60~\cite{Arn07} was based on the \sigmaabs\ value obtained from \mbox{p-A}
collisions at 400/450 GeV, assuming \sigmaabs\ to be independent of the incident
energy and $y_{\rm {CM}}$ range. As shown in Fig.\ref{fig:3} (left), this is not the case, since 
the cold nuclear matter effects strongly depend on the energy and kinematical 
domain. Therefore the evaluation of the size of the effects affecting
\mbox{A-A} collisions at 158A GeV must be based on the new \mbox{p-A} NA60 
results at the same energy.
The expected shape for the \jpsi\ distribution in cold nuclear matter is computed
within a Glauber model, 
assuming the \jpsi\ production as a function of centrality (determined measuring
the energy, $E_{ZDC}$, released in a Zero Degree Calorimeter) $dN^{exp}_{\rm J/\psi}/dE_{ZDC}$ to 
scale with the number of binary collisions. The nuclear 
effects affecting the \jpsi\ are then implemented assuming \sigmaabs\ = 7.6 $\pm$ 
0.7 $\pm$ 0.6 mb, as discussed in the previous section. This reference curve is then compared with
the measured \mbox{In-In} $dN^{meas}_{\rm J/\psi}/dE_{ZDC}$ distribution, following the procedure described
in~\cite{Arn07}.
As expected, since the \sigmaabs\ value directly obtained at 158 GeV is higher
than the one extracted from \mbox{p-A} data at 400/450 GeV, the
amount of anomalous suppression will be smaller with respect to the previous
estimates. 
Furthermore, in the procedure followed up to now to build the 
reference curve, shadowing effects were not explicitly taken into account.
While in \mbox{p-A} data only partons in the target are affected by shadowing,
in \mbox{A-A} collisions also the projectile is involved and this further
contribution has to be included when extrapolating the cold nuclear 
effects from \mbox{p-A} to \mbox{A-A}~\cite{Arn09}. 
It turns out that if the shadowing is neglected in this extrapolation, 
a small bias is introduced, resulting in a $\sim$5\% artificial 
contribution to the suppression of the \jpsi\ yield (if the EKS98
parameterization is used for the shadowing evaluation). 
Therefore, if shadowing is properly taken into account in the
extrapolation from \mbox{p-A} to \mbox{A-A}, a further 
reduction of the \jpsi\ suppression is observed. This is shown in 
Fig.\ref{fig:3} (right), where the final ratio of the measured and expected 
\jpsi\ yield is presented as a function of the number of participants 
($N_{part}$) for both \mbox{In-In} and \mbox{Pb-Pb} data.
Even within this new definition of the reference curve, for very central 
\mbox{Pb-Pb} collisions ($N_{part}>200$) an anomalous 
 \jpsi\ suppression, of the order of 20-30\%, is still visible.

\section{$\rm J/\psi$ polarization}\label{sec:pol}

The measurement of the \jpsi\ polarization is an important tool to clarify the
quarkonium production mechanism, since different theoretical models predict
different degrees of polarization~\cite{Lan08}. For example, polarization is 
expected to be sensitive to the spin states of the $c\bar{c}$ pair~\cite{Ben98}, 
therefore its measurement can shed some light on the color-singlet and color-octet 
contributions to the production process. 
Contrary to other observables, as the differential cross-sections, in the 
polarization predictions uncertainties related to the theoretical inputs cancel
out, therefore providing well-defined expected values.
However, quarkonium polarization has always 
represented a puzzle, since, at collider experiments, the theoretical models, which 
were able to predict the observed \jpsi\ production cross section, clearly 
failed in the description of the $p_{\rm T}$ dependence of the
polarization~\cite{Bra00}.

Experimentally, the \jpsi\ polarization is measured from the full angular 
distribution of the quarkonium decay products:
\begin{equation}\label{eq:1}
\centering
\frac{dN}{dcos\theta d\phi}=1+\lambda cos^{2}\theta + \mu sin2\theta cos\phi
+\frac{\nu}{2}sin^{2}\theta cos2\phi
\end{equation}
where the polar angle $\theta$ is determined by the 
direction of the positive muon, in the quarkonium rest frame, and a chosen
$\vec{z}$ polarization axis (defining the decay plane). $\phi$ is the azimuthal angle between the reaction plane,
containing the colliding hadrons and the decay plane. 
The $\lambda$ parameter is traditionally called ``polarization'': zero value
for $\lambda$ indicates that the \jpsi\ are
not polarized, while $\lambda$=1(-1) means transverse (longitudinal) polarization.
Non zero values of $\mu$ and $\nu$ indicate an azimuthal anisotropy of
the distributions.
$\lambda$, $\mu$ and $\nu$ depend on the chosen definition of the
$\vec{z}$ axis. It is, therefore, useful, 
to determine these parameters in more than one reference frame.
However, it can be noted that the knowledge of the full set
of coefficients allows to analytically calculate their values in other frames,
through appropriate transformations and knowing the \jpsi\ 
kinematics~\cite{Fal86}.
Of course, this is  
not the case if only the $\lambda$ parameter is measured.
 In the literature the most commonly used frames are the 
Collins-Soper (CS) one, where the $\vec{z}$ axis is parallel to the bisector of the angle
between the projectile and target directions in the \jpsi\ rest frame, and the
helicity (HE) one, where the $\vec{z}$ axis coincides with the quarkonium direction in the
target-projectile center of mass frame.
It is important to note that because of the $\lambda$, $\mu$ and $\nu$
dependence on a chosen reference frame, polarization results from different experiments 
can be compared only if the same frame is adopted. 
Most previous polarization analyses up to now were limited to the choice of only
one specific
reference frame and were restricted to the measurement of the
polar angle distribution.
On the contrary, recent results from HERA-B~\cite{Abt09_2} and NA60~\cite{Sco09}
have been obtained measuring for the first time the full angular distribution of
the decay products of charmonia and comparing different frames.

\begin{figure}[ht]
\centering
\includegraphics[scale=0.33]{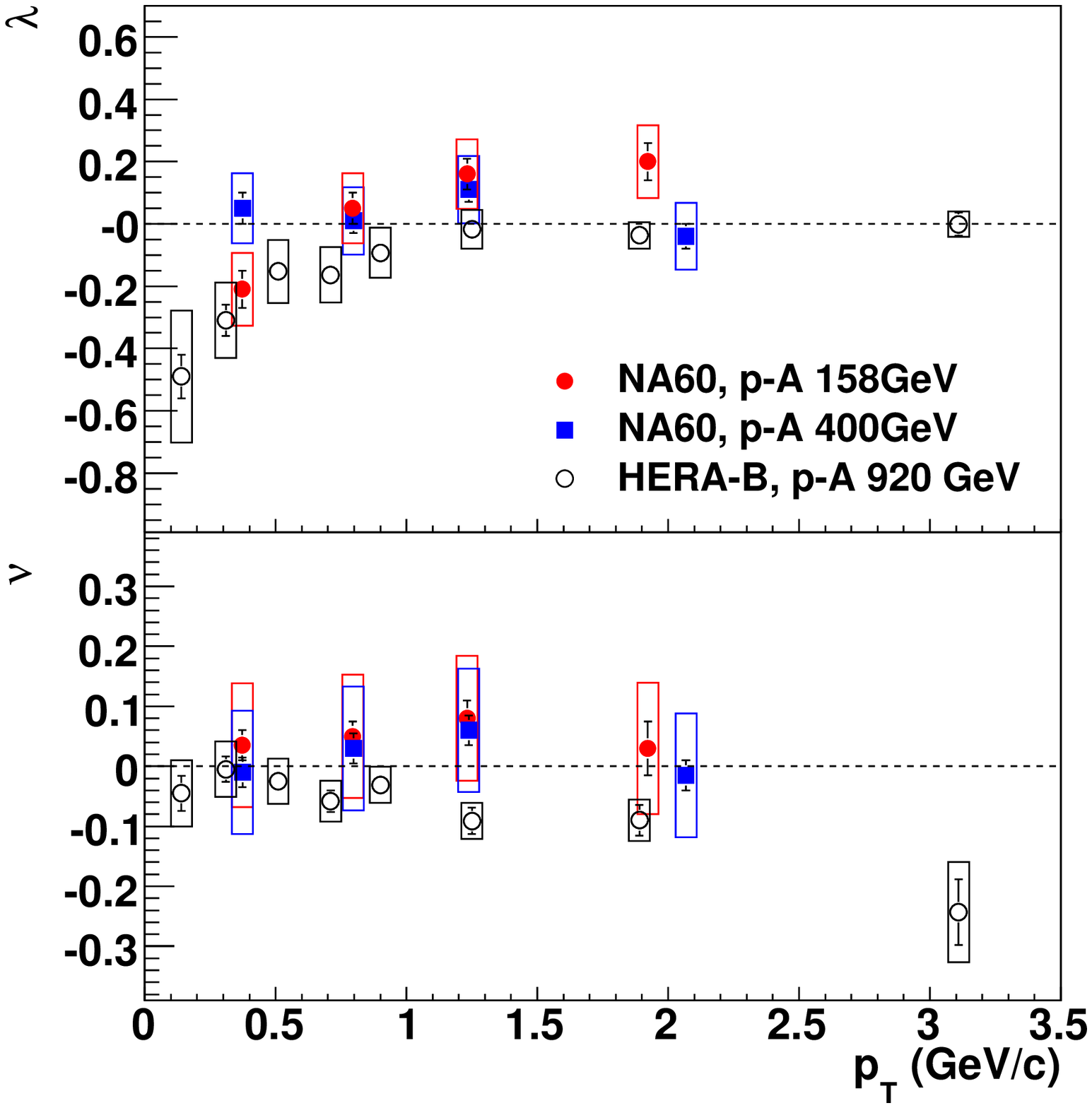}
\includegraphics[scale=0.33]{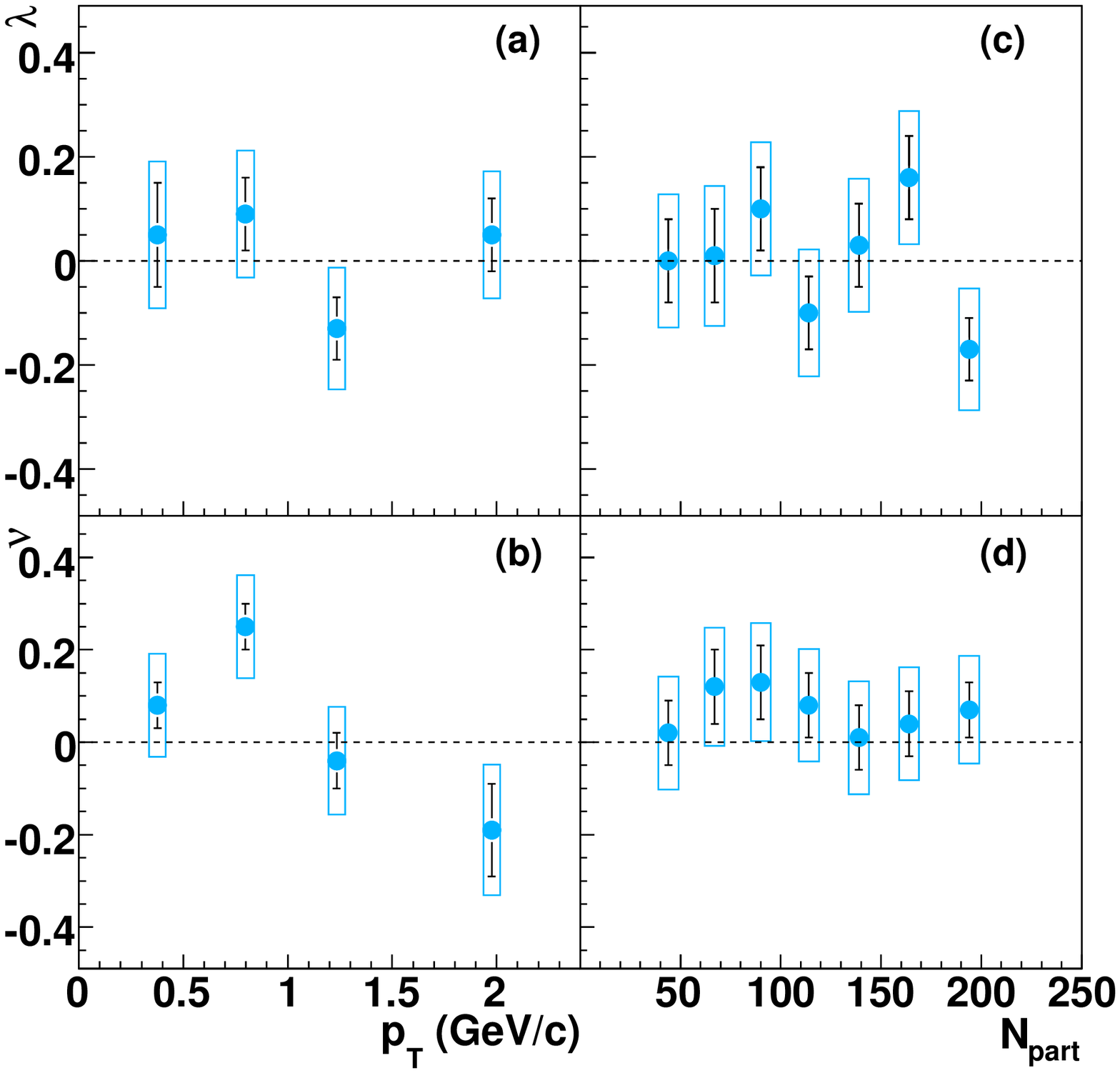}
\caption[]{Left: \jpsi\ polarization parameters $\lambda$ (top) and $\nu$ (bottom) 
as a function of $p_{T}$ in the helicity frame for \mbox{p-A} 
collisions from HERA-B and NA60. Because of a slighty different definition of
the $\nu$ parameter, with respect to HERA-B, the plotted NA60 $\nu$ values have
been divided by a factor 2 with respect to those obtained from Eq. \ref{eq:1}. Right: $\lambda$ (a,c) and $\nu$ (b,d) 
results for \mbox{In-In} collisions versus $p_{\rm T}$ and centrality respectively. 
The boxes around the points represent the total error
affecting the measurement, assuming, for the NA60 points, a conservative
preliminary 0.1 systematic error.}
\label{fig:4}
\end{figure}

NA60 results are obtained applying a 1-D acceptance correction, assuming
realistic $y_{\rm CM}$ and $p_{\rm T}$ distributions and a flat cos$\theta$ and
$\phi$ spectra.
The measured polarization values present a dependence on the \jpsi\ 
transverse momentum, as shown in Fig.\ref{fig:4} (left) for the helicity frame. 
In particular, HERA-B $\lambda$ results indicate an increase of the \jpsi\ 
polarization, moving from slightly negative values (corresponding to
longitudinal polarization) at low $p_{\rm T}$ to values
close to zero (absence of polarization) at higher $p_{\rm T}$. 
The first polarization results obtained in \mbox{p-A} collisions at
158 GeV and 400 GeV by the NA60 experiment confirm, within errors, the observed trend.

In Fig.\ref{fig:4} (left) results on the the azimuthal parameter $\nu$ are 
also shown, always in the helicity frame. In this
case no clear dependence on the \jpsi\ kinematics is visible.
Also the $\mu$ parameter is consistent with zero everywhere.
Although not shown here, both HERA-B and NA60 results in the 
Collins-Soper reference frame give $\lambda$ values which tend to be more 
negative
and larger in absolute value with respect to the ones measured in the helicity
frame.
This is an confirmation of the fact that the polarization parameters have a clear
dependence on the chosen frame.

The HERA-B and NA60 new polarization results clearly will shed 
some light on the polarization issue, in the attempt of describing all the
available polarization measurement in a common scenario.

NA60 has also provided, for the first time, results on the full angular
distribution of the \jpsi\ decay products in \mbox{In-In} collisions. 
The $\lambda$ values, in the helicity frame, do not show a dependence   
on the \jpsi\ $p_{\rm T}$ as shown in Fig\ref{fig:4} (a) and they
are almost consistent with zero everywhere. 
Also the azimuthal coefficient $\nu$ is rather
small, pointing to a positive value only for $p_{T} \sim 1$ GeV/c, as shown in
Fig\ref{fig:4} (b).
The $\lambda$ and $\nu$ values do not present any dependence even on the 
centrality of the collision, as shown in Fig.\ref{fig:4} (c,d). 
The pattern of these results is confirmed also if the Collins-Soper frame is adopted.
In principle, it may be expected that the formation of a hot medium may 
affect the \jpsi\ polarization, as proposed in ~\cite{Iof03}. Therefore,
quantitative predictions are needed, in order to clarify the observed results.

\section{Conclusions}

New NA60 results obtained in \mbox{p-A} collisions at 158 and 400 GeV have been
presented and compared with the already existing measurements from
other fixed target experiments. 
Cold nuclear matter effects, determined from \mbox{p-A} interactions, exhibit a 
rather strong energy and kinematical dependence. 
Therefore the use of the new \mbox{p-A} results at 158A GeV, the same energy as
the \mbox{A-A} data, allows a more precise definition
of the expected cold nuclear matter effects in nuclear collisions.
Apart from very central \mbox{Pb-Pb} collisions, where the anomalous suppression
is still sizeable, a smaller effect with respect to previous estimates is observed.
Results on \jpsi\ polar and azimuthal parameters have also been presented for 
both \mbox{p-A} and \mbox{A-A} collisions.



\end{document}